\documentclass[prl,amssymb,twocolumn,showpacs,superscriptaddress,floatfix]{revtex4-1}

\usepackage{graphicx}
\usepackage{bm}
\usepackage{mathptmx}
\usepackage[mathlines]{lineno}

\begin{document}

\title{Unconventional critical activated scaling of two-dimensional quantum spin glasses}

\author{D. A. Matoz-Fernandez}
\affiliation{Universit\'e Grenoble Alpes, LIPHY, F-38000 Grenoble, France}
\affiliation{CNRS, LIPHY, F-38000 Grenoble, France}
\author{F. Rom\'a}
\affiliation{Departamento de F\'{\i}sica, Universidad Nacional de
San Luis, INFAP, CONICET, Chacabuco 917, D5700BWS San Luis, Argentina}

\begin{abstract}
We study the critical behavior of two-dimensional short-range quantum spin glasses by numerical simulations.
Using a parallel tempering algorithm, we calculate the Binder cumulant 
for the Ising spin glass in a transverse magnetic field with two different 
short-range bond distributions, the bimodal and the Gaussian ones. 
Through an exhaustive finite-size analysis, we show that
the cumulant probably follows an unconventional activated scaling,
which we interpret as new evidence supporting the hypothesis that 
the quantum critical behavior is governed by an infinite randomness fixed point.   
\end{abstract}



\maketitle
Quantum phase transitions in condensed matter have been a subject of special interest 
though many decades \cite{Sachdev2011}.  
This phenomenon manifests itself in systems where quantum instead of thermal fluctuations 
are relevant.  An order-disorder phase transition can occur even at zero temperature,
if a suitable parameter (a magnetic field, for example) is tuned externally through the critical region.
Simple models, e. g. the pure Ising ferromagnet chain in a transverse field, 
have been used as prototypes for testing our understanding in the vicinity of such 
critical points \cite{Suzuki2013}.  More interesting still is the criticality found
in disordered systems.  It has been established that the quantum phase transition in 
diluted and random Ising models in a transverse field, is controlled by the so-called infinite 
randomness fixed point (IRFP) \cite{Fisher1999} which, among other things,
is characterized by a divergent dynamical exponent $z$ and an unconventional dynamic scaling 
\cite{Suzuki2013,Pich1998,Rieger2005}.    

The critical behavior of the quantum disordered and frustrated systems, however, 
is very poorly understood \cite{Sachdev2011}.  Spin glasses are the paradigmatic models 
of such theoretical challenge and, presumably, their phase transitions should govern 
by the IRFP \cite{Motrunich2000}. 
Although recent theoretical works \cite{Kovacs2010,Miyazaki2013,Monthus2015} support this conjecture, 
old Monte Carlo studies concluded that for two \cite{Rieger1994} and three \cite{Guo1994} dimensions, 
the quantum phase transition of such systems is instead conventional (with $z$ takes a finite value).
Subsequent simulation research has explored this same problem concluding that in 
two dimensions and at the critical point, several observables (different versions of the Binder cumulant 
and the correlation length) do not follow a conventional dynamic scaling \cite{Dayal}.   
Such disagreements are still an open question, which often is circumvented in favor of the IRFP scenario
by noting that small system sizes were used in these numerical works. 
Being that the simulations of disordered and highly frustrated systems as spin glasses
inevitably suffer from this drawback, at first sight this obstacle seems impossible to overcome
without the use of an alternative strategy. 

In this paper, we use a quantum parallel-tempering Monte Carlo algorithm 
to simulate the two-dimensional Ising spin glass model in a transverse magnetic field. 
Through an exhaustive finite-size scaling analysis of the Binder cumulants, 
we present new evidence for the existence of an IRFP in this system.      

The Hamiltonian of the two-dimensional Ising spin-glass model in a transverse magnetic field is
\begin{equation}
\mathcal{H}=-\sum_{\langle i,j\rangle}J_{ij}\sigma_{i}^{z}\sigma_{j}^{z}-\Gamma\sum_{i=1}^N \sigma_{i}^{x},
\label{Hamiltonian}
\end{equation}
where the first sum runs over the pairs of nearest-neighbor sites of a square lattice of linear size $L$ (with $N=L^2$ spins), $\sigma_i$ are Pauli spin matrices, $\Gamma$ is the strength of the transverse field, and the interactions $J_{ij}$ are independent random variables drawn from a given distribution with mean zero and variance one.  We consider both, the bimodal ($\pm 1$) and the Gaussian bond distributions.

To perform a Monte Carlo simulation, first we use the Suzuki-Trotter formalism \cite{Suzuki-Trotter} to map the $d$-dimensional quantum model onto an effective $(d+1)$-dimensional classical one, whose action is \cite{Rieger1994}        
\begin{equation}
\label{Action}
\mathcal{A}=-\sum_{\tau=1}^{L_\tau} \sum_{\langle i,j\rangle} K_{ij} S_i(\tau) S_j(\tau)
-K \sum_{\tau=1}^{L_\tau} \sum_{i=1}^N S_i(\tau) S_i(\tau+1),
\end{equation}
where $K_{ij}=\Delta \tau J_{ij}$ and $K=\frac{1}{2}\ln [\coth\left(\Delta \tau \Gamma\right)]$, $S_i=\pm 1$ are classical Ising spins, and the index $i$\,($j$) run over the sites of the original square lattice. Here $\tau$ represent the imaginary time or Trotter-dimension, 
which we divide into $L_\tau$ slices of width $\Delta \tau=1/T L_\tau$, 
with $T$ being the temperature. 
To strictly reproduce the ground state of the quantum Hamiltonian Eq.~(\ref{Hamiltonian}), we need take $\Delta \tau\rightarrow 0$.
However, as it has been argued elsewhere \cite{Rieger1994,Guo1994}, the universal properties of the phase transition should not depend on the short-length-scale details of the model, and therefore we can take $\Delta \tau=1$ without any loss of generality. Then, by setting the standard deviation of $K_{ij}$ equal to $K$, the Hamiltonian of the $(d+1)$-dimensional system is written as 
\begin{equation}
\label{ClassicalHamiltonian}
\mathcal{H}_\mathrm{cl}=-\sum_{\tau=1}^{L_\tau} \sum_{\langle i,j\rangle} J_{ij} S_i(\tau) S_j(\tau)
- \sum_{\tau=1}^{L_\tau} \sum_{i=1}^N S_i(\tau) S_i(\tau+1),
\end{equation}
with $K^{-1}$ acting as an effective temperature for the classical model.  
Thus, the statistical weight of each spin configuration 
is proportional to $\exp(-K \mathcal{H}_\mathrm{cl})$.

We simulate the classical model (\ref{ClassicalHamiltonian}) using a Monte Carlo parallel-tempering algorithm \cite{Hukushima1996}, with $12$ replicas of the system set at temperatures between $K_i^{-1}=3.3$ and $K_f^{-1}=3.6$ ($K_i^{-1}=3.2$ and $K_f^{-1}=3.4$) for the bimodal (Gaussian) case. The calculations were carried out for cubic lattices of size $L \times L \times L_\tau$ with fully periodic boundary conditions, and the largest system reached was $20\times 20\times 96$ for which $10^4$ Monte Carlo sweeps were necessary to achieve equilibrium. All quantities were averaged over $6 \times 10^3 $ different disorder samples. In particular, for the Gaussian case, it was necessary to simulate a set of systems of larger sizes up to $24\times 24\times 96$. 

We focus on the Binder cumulant \cite{BinderBook}
\begin{equation}
g_\mathrm{av}=\frac{1}{2}\left[3-\frac{\left\langle q^4 \right\rangle}{\left\langle q^2 \right\rangle^2}\right]_{\mathrm{av}}, \label{cumulant}
\end{equation}
where $\left\langle \cdots \right\rangle$ and $\left[\cdots\right]_{\mathrm{av}}$ denote thermal and disorder averages, respectively.  $q$ is the Edward-Anderson order parameter which is defined by the overlap between the configurations of two replicas of the system, $\alpha$ and $\beta$, with the same disorder,
\begin{equation}
\label{Eq. 3}
q=\frac{1}{L^2 L_{\mathrm{\tau}}}\sum_{i,\tau}S_i^\alpha(\tau) \ S_j^\beta(\tau).
\end{equation}  
If the dynamical exponent $z$ is finite, the Binder cumulant (\ref{cumulant}) is expected to obey the conventional finite-size scaling form 
\begin{equation}
g_\mathrm{av}=\tilde{g}_\mathrm{c} \left(\delta L^{1/\nu},L_\tau/L^z\right).
\label{CS1}
\end{equation}
Here $\delta=K/K_c-1$, with $K_c^{-1}$ being the critical temperature, is the distance from the critical point, and $\nu$ is the exponent for the average correlation length \cite{Rieger2005}. On the other hand, within an IRFP scenario, the cumulant should follow an unconventional finite-size scaling 
\begin{equation}
g_\mathrm{av}=\tilde{g}_\mathrm{u} \left(\delta L^{1/\nu},\ln L_\tau/L^\psi\right),
\label{US1}
\end{equation}
where $\psi$ is called the activated exponent \cite{Pich1998}. To determine which of these scaling relationships is the correct one, we need to perform a comprehensive study of the Monte Carlo data.

\begin{figure}[t!]
\includegraphics[width=\linewidth,clip=true]{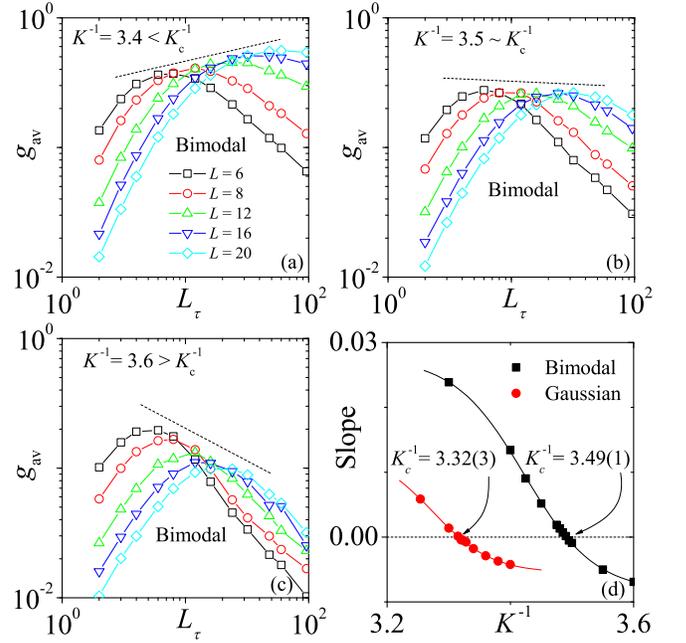}
\caption{\label{figure1} (a)-(c) Show the Binder cumulant for the bimodal case, as function of $L_\tau$ for different lattice sizes $L$ and three temperatures as indicated. (d) Shows for both, the bimodal and the Gaussian cases, the slope of the straight line that intersects the maxima of the Binder ratio, $g_{\mathrm{av}}^{\mathrm{max}}$, against $K^{-1}$.}
\end{figure}

First of all, we calculate the critical temperature following the lines of Refs.\cite{Rieger1994,Guo1994}. Because the Binder cumulant vanishes for a disordered phase, it is expected that when $L \to \infty$ for fixed $L_\tau$, as well as when $L_\tau \to \infty$ for fixed $L$, $g_\mathrm{av} \to 0$.  The reason is simple: In the first limit the model tends to a classical two-dimensional spin glass, while in the second limit it turns into an effective one-dimensional ferromagnetic chain, both systems having a disordered phase at any finite temperature. In between these extremes the Binder cumulant
reaches a maximum, making evident the existence of an ordered phase. Besides, both scaling relations (\ref{CS1}) and (\ref{US1}) predict that
at the critical temperature ($\delta=0$) and if a suitable relation between $L$ and $L_\tau$ is imposed (since the system is very anisotropic), this maximum does not depend on $L$.     

This last observation suggests a simple way to determine $K_c^{-1}$. Figures~\ref{figure1} (a)-(c) show, for bimodal interactions, the Binder cumulant 
as a function of $L_\tau$ for different lattice sizes $L$ and, respectively, for temperatures $K^{-1}<K_c^{-1}$, $K^{-1} \approx K_c^{-1}$, and $K^{-1}>K_c^{-1}$. In each cases, the maximum values of the Binder ratio, $g_{\mathrm{av}}^{\mathrm{max}}$, describes approximately a straight line whose slope vanishes at the critical point. Then, by plotting this slope against $K^{-1}$, Fig.~\ref{figure1} (d), we can calculate a very accurate value for the critical temperatures.  We obtain $K_c^{-1}=3.49(1)$ for the bimodal case. To our knowledge this critical temperature had not been previously calculated. On the other hand, for the Gaussian case we obtain $K_c^{-1}=3.32(3)$, a value very close to that reported by Rieger and Young, $K_c^{-1}=3.275(25)$ \cite{Rieger1994}.

\begin{figure}[t!]
\includegraphics[width=7.5cm,clip=true]{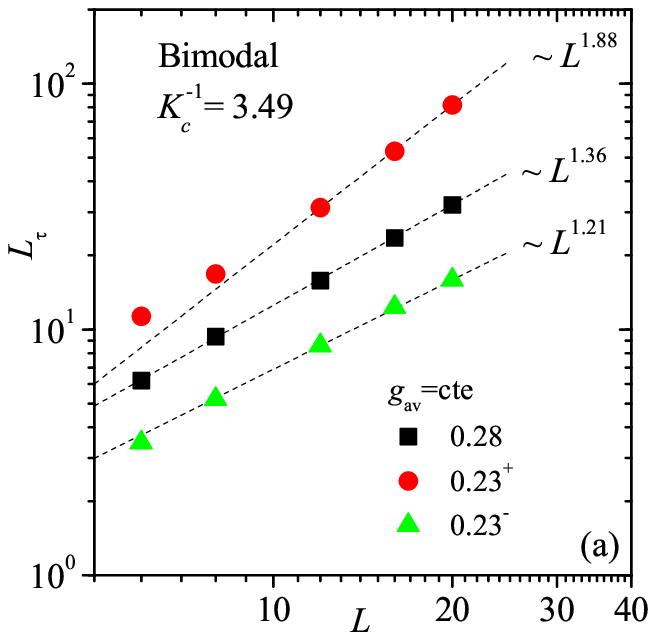}
\includegraphics[width=7.5cm,clip=true]{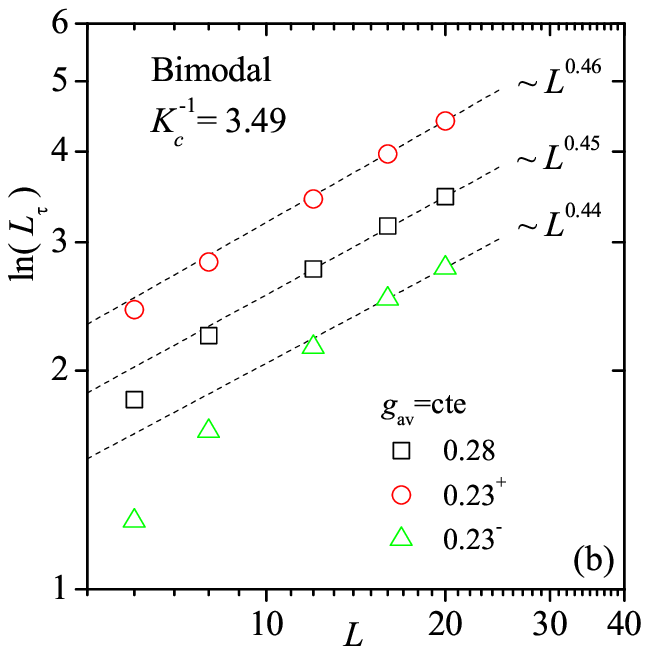}
\caption{\label{figure2} The dependence of (a) $L_\tau$ and 
(b) $\ln (L_\tau)$ with $L$, for different values of $g_\mathrm{av}$ as indicated.  
Curves are plotted in a log-log scale.}
\end{figure}

Having found the critical points we carry out, for each system, new simulations at exactly the corresponding critical temperatures 
[the curves at $K_c^{-1}$ look like that displayed in Fig.~\ref{figure1} (b)].
Then, the data set obtained is analyzed in the light of the scaling relations (\ref{CS1}) and (\ref{US1}). A simple way to decide which of these two functions is the right one, consist in plotting $L_\tau$ versus $L$ for constant $g_\mathrm{av}$.  According to Eq.~(\ref{CS1}), at the critical point ($\delta=0$) these lengths should be related as $L_\tau \sim L^z$. In the bimodal case Fig.~\ref{figure2} (a) shows that, for the maximum ($g_{\mathrm{av}}^{\mathrm{max}} \approx 0.28$), this scaling is met very well with $z \approx 1.36$.  However, for the Gaussian case, although we observe a similar behavior the exponent obtained is $z \approx 1.5$, a little different but compatible with the value previously calculated in Ref.~\cite{Rieger1994}. On the other hand, according to Eq.~(\ref{US1}), the true relation between $L_\tau$ and $L$ should be $\ln (L_\tau) \sim L^\psi$. Figure~\ref{figure2} (b) seems to indicate that, for the maximum of the Binder ratio, this functionality is probably only fulfilled for large lattice sizes with an exponent $\psi \approx 0.45$. Also, for Gaussian interactions, we observe a similar trend with $\psi \approx 0.46$.  

\begin{figure}[t!]
\includegraphics[width=7.5cm,clip=true]{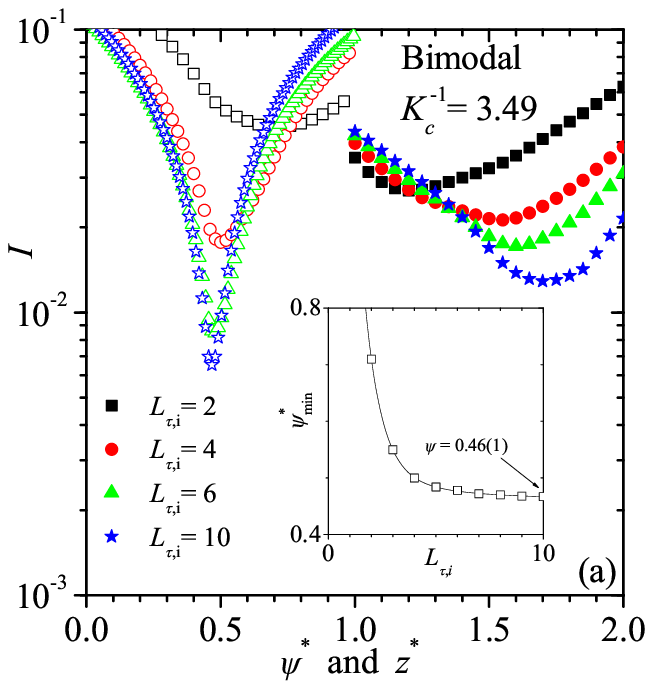}
\includegraphics[width=7.5cm,clip=true]{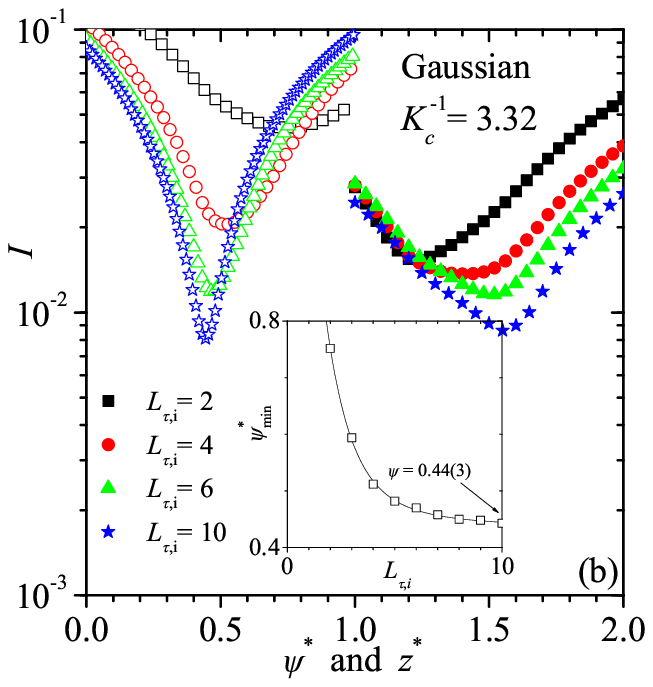}
\caption{\label{figure3} $I(z^*)$ (solid symbols)
and $I(\psi^*)$ (open symbols) for different $L_{\tau,i}$ as indicated, for 
(a) the bimodal and (b) the Gaussian systems at the critical point.  
The insets show how $\psi^*_\mathrm{min}$ depends on $L_{\tau,i}$ (see text).}
\end{figure}

For other values of $g_\mathrm{av}$, Fig.~\ref{figure2} (a) shows that the conventional scaling fails because different values of $z$ should be considered to fit the data well.  Here, $g_\mathrm{av}=0.23^-$ ($g_\mathrm{av}=0.23^+$) correspond to points with $g_\mathrm{av}=0.23$ but that lies to the left (right) of the maximum. This drawback does not occur for the unconventional scaling [see Fig.~\ref{figure2} (b)], since a single value of
$\psi$ is sufficient to describe approximately the data range.  The same is observed for the Gaussian case and also using $\psi \approx 0.46$. 
In this context we see that the hypothesis, assumed by us above, that the universality class does not depend on the exact form of the bond distribution, is valid only if the unconventional scaling is the correct one. 

A more comprehensive study can be done by performing a data collapse analysis, thereby determining the best candidate values for the critical exponents $z$ and $\psi$. Specifically, to test the scaling relation (\ref{CS1}) at the critical point, we plot the Binder cumulant for all lattice sizes as function of $L_\tau / L^{z^*}$ and, for different values of $z^*$, we calculate a suitable function $I(z^*)$ in order to measure the goodness of the collapse.  We choose $I(z^*)$ equal to the normalized sum of the areas between all pairs of curves that are contiguous in $L$, \textit{i.e.}, those for which the difference between the corresponding lattice sizes is the smallest (namely, $L=6$ with $L=8$, $L=8$ with $L=12$, etc). Then, the best candidate value for $z$, $z^*_\mathrm{min}$, is obtained by minimizing this special function. Furthermore, to analyze the unconventional scaling we proceed in the same way, but now we plot the Binder cumulant as a function of $\ln (L_\tau) / L^{\psi^*}$, and then we minimize $I(\psi^*)$ to calculate $\psi^*_\mathrm{min}$. The details of this procedure are given in the Supplemental Material \cite{Supplemental}. 

For the bimodal case, Fig.~\ref{figure3} (a) shows what happens  
when we calculate the function $I$ using all data available.  
That is, by doing the calculations taking into account systems 
with $6 \le L \le 20$ and $2\le L_\tau \le 96$. 
These curves are labeled with $L_{\tau,i}=2$, the smallest value of $L_\tau$ in the set. 
From the conventional scaling (solid black squares) 
we obtain a minimum at $z^*_\mathrm{min} \approx 1.21$, 
while for the unconventional one (open black squares) 
this extreme is located at $\psi^*_\mathrm{min} \approx 0.71$, 
the former being the deepest. A direct interpretation of 
this result tells us that the best data collapse is achieved 
within the conventional framework. However, this is a hasty conclusion.  

By simple inspection of the procedure used, it is easy to see 
that the Binder cumulants of systems with the smaller sizes 
dominate such calculations. Then, to overcome finite-size effects, 
we calculate again the function $I$ but now gradually removing 
such small lattices starting from low to high values of 
$L_{\tau,i}$, \textit{i.e.}, considering only systems 
with $6 \le L \le 20$ and $L_{\tau,i}\le L_\tau \le 96$. 
Figure~\ref{figure3} (a) shows also the curves for $L_{\tau,i}=4$, 6, and 10. 
From these plots arise two important observations: 
The minimum of $I$ for the unconventional scaling is always the deepest and, 
more important, $\psi^*_\mathrm{min}$ converges quickly to $\psi=0.46(1)$ 
[see inset in Fig.~\ref{figure3} (a)] while $z^*_\mathrm{min}$ changes 
continuously without apparently reaching a limit value 
(at least for $L_{\tau,i}=10$, $z^*_\mathrm{min} \approx 1.7$). 

For Gaussian interactions the finite-size effects are larger. 
To overcome this problem, we simulate systems of dimensions 
up to $24\times 24\times 96$ increasing our data set to 
$6 \le L \le 24$ and $2\le L_\tau \le 96$. 
Figure~\ref{figure3} (b) shows the functions $I(z^*)$ 
and $I(\psi^*)$ for $L_{\tau,i}=2-10$. 
The data show the same trend observed for the bimodal case,
but now the convergence is much slower: 
$\psi^*_\mathrm{min}$ converges to $\psi=0.44(3)$, 
while $z^*_\mathrm{min}$ does not tend to a definite limit.
Nevertheless, $z^*_\mathrm{min} \approx 1.55$ for $L_{\tau,i}=10$.

These results suggest again that, 
for the range of system sizes studied here,
the unconventional scaling is the most appropriate to achieve 
a consistent data collapse of the Binder cumulants.  

\begin{figure}[t!]
\includegraphics[width=7.5cm,clip=true]{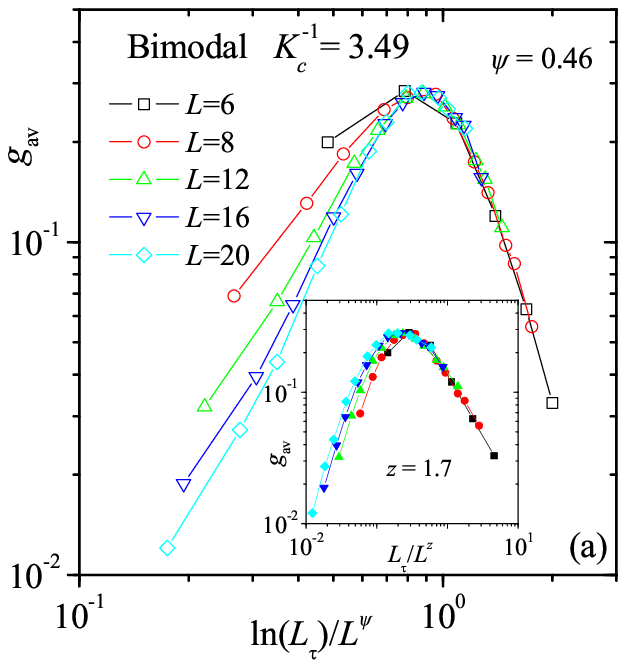}
\includegraphics[width=7.5cm,clip=true]{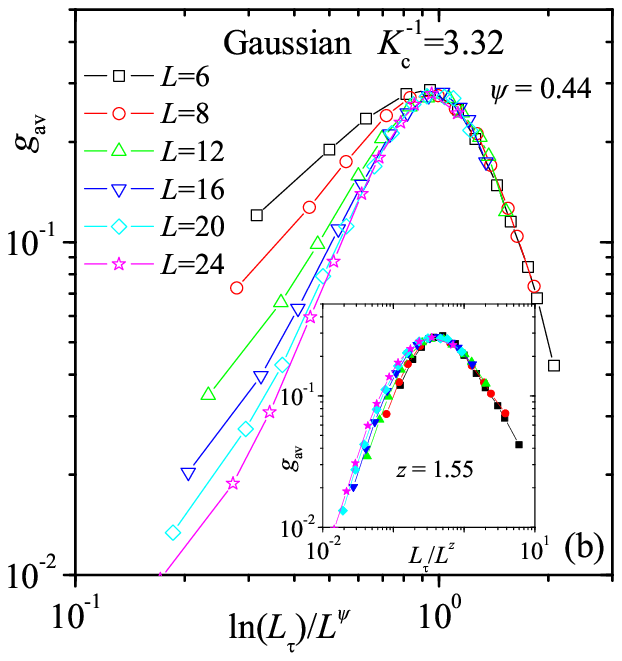}
\caption{\label{figure4} (Color online) The unconventional data collapse of the Binder cumulants for (a) the bimodal and (b) the Gaussian systems.  The insets show the respective conventional data collapses.}
\end{figure}

Finally, Figs.~\ref{figure4} (a) and \ref{figure4}(b) show, respectively, 
the unconventional data collapse of the Binder cumulants at 
the critical point for the bimodal and the Gaussian systems, 
where we have used the above calculated values of $\psi$. 
For each case inset also shows the (best) conventional data collapse. 
At first sight we observe that, in contradiction with our previous findings, 
the latter looks like the most adequate scaling because the corresponding curves overlap nicely, 
while the points to the left of the peak for the unconventional one 
does not collapse completely well. 
Notice, however, that these points correspond 
to the smaller values of $L_\tau$ discarded in our calculations
in order to overcome finite-size effects. 
This shows that a ``qualitative'' analysis looking for good data collapses,
is not enough to replace the ``quantitative'' and systematic 
procedure presented in this work.  

In summary, we have carried out an exhaustive scaling 
analysis of the Binder cumulant for a two-dimensional quantum spin glass 
in a transverse magnetic field with both, bimodal and Gaussian interactions. 
We determine that, at the critical point, the most probable scenario 
is that such a data set follows an unconventional finite-size scaling 
(\ref{US1}) with an activated exponent $\psi \simeq 0.44 - 0.46$.  
These values are compatible with $\psi = 0.48(2)$ obtained by a strong 
disorder renormalization group method \cite{Kovacs2010}, 
but are very different from $\psi \simeq 0.65$ calculated recently by block renormalization \cite{Monthus2015}.  
In addition, from the derivate of $g_\mathrm{av}$ with respect to $K$ 
at the critical point, we have also calculated $\nu=1.2(4)$ (bimodal) and $\nu=1.13(5)$ (Gaussian), 
the exponents for the average correlation length.  These values agree very well  
with those obtained previously: $\nu = 1.24(2)$ \cite{Kovacs2010}, 
$\nu=1.21(9)$ \cite{Miyazaki2013}, and $\nu \simeq 1.25$ \cite{Monthus2015}. 

In conclusion, our findings support the hypothesis that the critical 
behavior of this two-dimensional quantum spin glass model is controlled by an IRFP,
a result contrary to the standard picture reported in Ref.~\cite{Rieger1994}, 
probably the only available simulation study of such a system.

This work was supported in part by CONICET under Project No. PIP 112-201301-00049, 
by FONCyT under Project No. PICT-2013-0214, and by Universidad Nacional de San Luis under 
Project No. PROIPRO 3-10214 (Argentina). We thank to LIPhy-UJF (France) for computational resources.


\newpage

\vspace{2cm}

\begin{widetext}
\begin{center}
{\large
Supplementary Material for:\\\vspace{2mm}
\vspace{0.2cm}
\textbf{Numerical evidence for an unconventional critical behavior in two-dimensional quantum spin-glasses}\\\vspace{2mm}
\vspace{0.2cm}
D. A. Matoz-Fernandez and F. Rom\'a}
\end{center}
\end{widetext}

\vspace{0.5cm}

\begin{figure}[t!]
\includegraphics[width=6cm,clip=true]{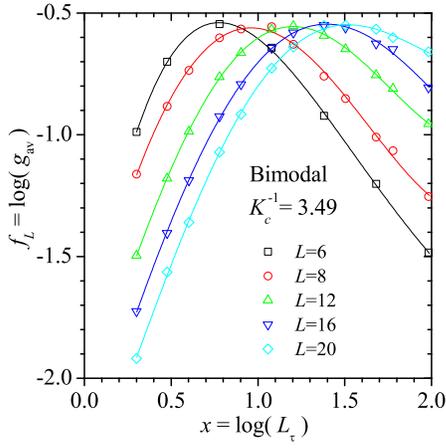}
\caption{\label{figure1supp} (Color online) The logarithm of the Binder cumulant versus the logarithm of $L_\tau$, for the bimodal case and for different lattice sizes $L$ as indicated.  Open points correspond to simulation results while continuous lines are fits made with a fourth-order polynomial.}
\end{figure}

The data collapse analysis is explained taking as an example the bimodal case and, in particular, the unconventional scaling. Figure~\ref{figure1supp} shows the logarithm of the Binder cumulants, $f_L=\log(g_{\mathrm{av}})$, as function of $x=\log(L_\tau)$.  The Monte Carlo data is represented by open points for lattices with $6 \le L \le 20$ and $2\le L_\tau \le 96$.

\begin{figure}[b!]
\includegraphics[width=6cm,clip=true]{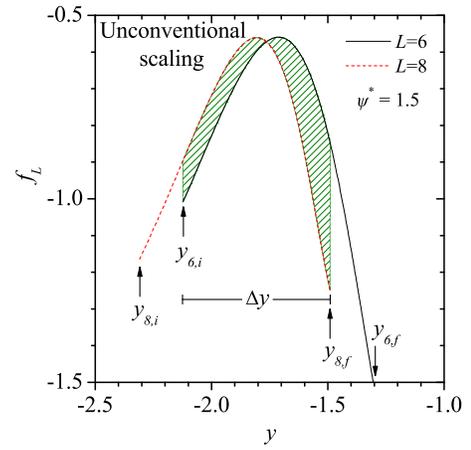}
\caption{\label{figure2supp} (Color online) Unconventional data plot of $f_L$ versus $y$, for lattice sizes $L=6$ (continuous black line) and $L=8$ (dashed red line) and for $\psi^*=1.5$.  The area between both curves (green pattern) is limited to a coincidence range of width $\Delta y$. }
\end{figure}

In order to analyze this data set, first we fit each curve with a fourth-order polynomial 
\begin{equation}  
f_L(x)=A_L + B_L \ x+ C_L \ x^2 + D_L \ x^3 + E_L \ x^4,  \label{pol} 
\end{equation}
where the coefficients $A_L-E_L$ depend on $L$.  
Continuous lines in Fig.~\ref{figure1supp} correspond to such fits. From now on, we work exclusively with these continuous functions. To try an unconventional data collapse for the exponent $\psi^*$, we need to plot $f_L$ as function of
\begin{equation}   
y=\log[\ln(L_\tau)/L^{\psi^*}]= \log(x)-\log[\log(e)]-\psi^* \log(L),        
\end{equation}
where $e$ is the Euler number.  Function $f_L(y)$ is given by the polynomial (\ref{pol}) replacing $x$ by $x=10^y L^{\psi^*} \log(e)$. 
Figure~\ref{figure2supp} shows an example for two lattice sizes, $L=6$ and 8, and for $\psi^*=1.5$.  For each curve, variable $y$ range between \begin{equation}   
y_{L,i}=\log[\log(L_{\tau,i})]-\log[\log(e)]-\psi^* \log(L) 
\label{yi}
\end{equation}
and 
\begin{equation}   
y_{L,f}=\log[\log(L_{\tau,f})]-\log[\log(e)]-\psi^* \log(L),
\label{yf}
\end{equation}
with $L_{\tau,i}=2$ (the smallest value of $L_\tau$)  and $L_{\tau,f}=96$ (the largest value of $L_\tau$) for the full data set. Because the extremes (\ref{yi}) and (\ref{yf}) depend on $\psi^*$, for a given value of this exponent the coincidence range of a pair of curves of sizes $L_a$ and $L_b$ is limited to 
\begin{equation}   
y_\mathrm{min}=\mathrm{max}\{y_{L_a,i},y_{L_b,i}\} 
\end{equation}
and
\begin{equation}   
y_\mathrm{max}=\mathrm{min}\{y_{L_a,f},y_{L_b,f}\},
\end{equation}
being $\Delta y=y_\mathrm{max}-y_\mathrm{min}$ (see Fig.~\ref{figure2supp}).

To quantify how good is a given exponent value, we calculate the following integral 
\begin{equation}
I_{a,b}=\frac{1}{\Delta y} \int_{y_\mathrm{min}}^{y_\mathrm{max}} \Big|f_{L_a}-f_{L_b} \Big| \ dy.
\label{Iab}
\end{equation}
This is a function of $\psi^*$ which measures the area difference between the curves and it is normalized by $\Delta y$.  The normalization is chosen 
so as to allow for comparison between the results derived from the unconventional and the conventional data collapses. Finally, in order to calculate $I$, we average $I_{a,b}$ over all pairs of curves with sizes $(L_a,L_b)$ that are contiguous in $L$ [namely, $(6,8)$, $(8,12)$, $(12,16)$, and $(16,20)$]
\begin{equation} 
I=\frac{1}{P} \sum_{\mathrm{pares}} I_{a,b} \ ,
\end{equation}
where $P=4$ is the total number of pairs.

On the other hand, to make a conventional data collapse for the exponent $z^*$,
we proceed in a similar way.  Namely, we plot $f_L$ as function of
\begin{equation}   
y=\log(L_\tau/L^{z^*})=x-z^*\log(L),        
\end{equation}
where now $f_L(y)$ is given by the polynomial (\ref{pol})
replacing $x$ by $x=y+z^*\log(L)$, and for each curve we calculate
the range of variable $y$, from
\begin{equation}   
y_{L,i}=\log(L_{\tau,i})-z^*\log(L) 
\end{equation}
to
\begin{equation}   
y_{L,f}=\log(L_{\tau,f})-z^*\log(L). 
\end{equation}
The rest of the procedure is essentially the same as before.

In addition, for both the conventional and the unconventional scalings, we have tried other forms for function $I$ and also we have made fits to higher-order polynomials, but our findings do not change appreciably. 

\begin{thebibliography}{99}
\bibitem{Sachdev2011} S. Sachdev, {\it Quantum Phase Transitions} (Cambridge University Press, Cambridge, U.K., 2011).
\bibitem{Suzuki2013} S. Suzuki, J.-I. Inoue, and B. K. Chkarabarti, {\it Quantum Ising Phases and Transitions in Transverse Ising Models} Lecture Notes in Physics, Vol. 862 (Springer, New York, 2013).
\bibitem{Fisher1999} D. S. Fisher, Physica A {\bf 263}, 222 (1999).
\bibitem{Pich1998} C. Pich, A. P. Young, H. Rieger, and N. Kawashima, Phys. Rev. Lett. {\bf 81}, 5916 (1998).
\bibitem{Rieger2005} H. Rieger, in {\it Quantum Annealing and Related Optimization Methods}, edited by A. Das and B. K. Chakrabarti (Springer Verlag, Berlin, 2005).
\bibitem{Motrunich2000} O. Motrunich, S.-C. Mau, D. A. Huse, and D. Fisher, Phys. Rev. B {\bf 61}, 1160 (2000).
\bibitem{Kovacs2010} I. A. Kov\'acs and F. Igl\'oi, Phys. Rev. B {\bf 82}, 054437 (2010).
\bibitem{Miyazaki2013} R. Miyazaki and H. Nishimori, Phys. Rev. E {\bf 87}, 032154 (2013).
\bibitem{Monthus2015} C. Monthus, J. Stat. Mech.: Theor. Exp. P01023 (2015).
\bibitem{Rieger1994} H. Rieger and A. P. Young, Phys. Rev. Lett. {\bf 72}, 4141 (1994).
\bibitem{Guo1994} M. Guo, R. N. Bhatt, and D. A. Huse, Phys. Rev. Lett. {\bf 72}, 4137 (1994).
\bibitem{Dayal} P. Dayal, PhD Thesis,  Swiss Federal Institute of Technology of Zurich (2006); http://dx.doi.org/10.3929/ethz-a-005244156.
\bibitem{Suzuki-Trotter} H. F. Trotter, Proc. Am. Math. Soc. {\bf 10}, 545 (1959); 
M. Suzuki, Progr. Theor. Phys. {\bf 56}, 1454 (1976).
\bibitem{Hukushima1996} K. Hukushima, K. Nemoto, J. Phys. Soc. Jpn. {\bf 65}, 1604 (1996).
\bibitem{BinderBook} K. Binder, {\it Applications of the Monte Carlo Method in Statistical
Physics. Topics in current Physics}, Vol. 36 (Springer, Berlin, 1984).
\bibitem{Supplemental} See Supplemental Material at http://link.aps.org/supplemental/
10.1103/PhysRevB.94.024201 for the data collapse analysis used to calculate the critical exponents.
\end{thebibliography}
\end{document}